\documentstyle[twocolumn,aps,prl,epsf]{revtex}
\begin{document}

\twocolumn[\hsize\textwidth\columnwidth\hsize\csname
@twocolumnfalse\endcsname

\title{Elastic Properties of C and $\mbox{B}_x\mbox{C}_y\mbox{N}_z$ 
       composite nanotubes}

\author{E.~Hern\'{a}ndez$^1$\thanks{To whom correspondence should be addressed.
E-mail address: ehe\@boltzmann.fam.cie.uva.es}, C.~Goze$^2$ and A.~Rubio$^1$} 
\address{$^1$~Departamento de F\'{\i}sica Te\'{o}rica, 
	Universidad de Valladolid, 
	Valladolid 47011, Spain \\
        $^2$~Groupe de Dynamique des Phases Condens\'{e}es, 
         Universit\'{e} Montpellier II, 34090 Montpellier, 
	France}

\maketitle

\begin{abstract}
We present a comparative study of the energetic, structural and elastic
properties of carbon and composite single-wall nanotubes, including
BN, $\mbox{BC}_3$ and $\mbox{BC}_2\mbox{N}$ nanotubes, using a
non-orthogonal tight-binding formalism. 
Our calculations predict that carbon nanotubes have a higher Young modulus 
than any of the studied composite
nanotubes, and of the same order as that found for defect-free graphene sheets. 
We obtain good agreement with the available experimental results.

\end{abstract}

]

\narrowtext

Carbon nanotubes~\cite{nanotube_reviews} were first discovered by 
Iijima~\cite{iijima} in the early nineties,
as a by-product of fullerene synthesis. Since then there has been an 
ever-increasing interest in these new forms of carbon, partly due to their
novel structures and properties, but perhaps more so due to the wealth of
potentially important applications in which nanotubes could be used. Indeed many
applications have already been reported, from their use as atomic-force
microscope tips~\cite{stm}, 
to field emitters~\cite{field}, 
nanoscale electronic devices~\cite{devices} or hydrogen 
storage~\cite{hydrogen}, to cite a few. But probably the highest potential
of nanotubes is in connection with their exceptional mechanical properties. 

After the discovery of graphitic nanotubes it was postulated that other
compounds forming laminar graphite-like structures could also form 
nanotubes~\cite{composite:reviews}.
In particular, BN~\cite{rubio:corkill:cohen}, $\mbox{BC}_3$~\cite{miyamoto},
$\mbox{BC}_2\mbox{N}$~\cite{BCN} and CN~\cite{CN} nanotubes were predicted 
on the basis of theoretical calculations. BN~\cite{bn_synthesis},  
$\mbox{BC}_3$ and $\mbox{BC}_2\mbox{N}$~\cite{bc3_synthesis} have since been 
synthesized,
though some uncertainty remains as to the actual structure of 
$\mbox{BC}_2\mbox{N}$ nanotubes~\cite{segregation}.

The mechanical properties of carbon nanotubes have been the subject of 
a number of 
theoretical~\cite{robertson,ruoff,molina,yakobson,cornwell,iijima:bernholc,lu}
as well as experimental~\cite{treacy,lieber,zettl} studies. On the theoretical
side, studies have been mostly carried out using empirical potentials, 
although Molina {\em et al.\/}~\cite{molina} employed an
orthogonal tight-binding model in their work. The most extensive theoretical
study of the elastic properties of carbon nanotubes to date is that of 
Lu~\cite{lu}, who used an empirical pair potential model to estimate the
Young modulus, Poisson ratio and other elastic constants of both single-wall
and multi-wall nanotubes, as well as nanotube ropes. However, it was
not possible to extend this study to composite nanotubes, given that no
empirical potential models akin to that used for carbon exist for
the composite systems. The behavior of carbon nanotubes subject to large 
axial strains has been studied by Yakobson {\em et al.\/}~\cite{yakobson}. 
The bending of carbon nanotubes has been studied experimentally and 
using simulation techniques by Iijima {\em et al.\/}~\cite{iijima:bernholc}.
The Young modulus of carbon multi-wall nanotubes has been experimentally
determined by Treacy {\em et al.\/}~\cite{treacy} using thermal 
vibration analysis of cantilevered tubes. They obtained a mean value of the
Young modulus of $1.8\pm 1.4$~TPa.
More recently, Wong {\em et al.\/}~\cite{lieber}
have obtained a value of $1.28\pm 0.59$~TPa, by recording the force needed to 
bend anchored nanotubes using an AFM. Chopra and Zettl~\cite{zettl}
have also used thermal vibration analysis to estimate the Young modulus of
multi-wall BN nanotubes, obtaining a value of $1.22\pm 0.24$~TPa. These 
experimental and theoretical studies confirm the expectation that nanotubes 
have exceptional stiffness, and could therefore 
be used in the synthesis of highly resistant composite materials.

In this work we study the structural, energetic and mechanical properties 
of both carbon and composite nanotubes, paying special attention to the 
mechanical properties, since these are expected to play such an important role
in many practical applications. This is the first time that such a detailed
comparative study is undertaken.
In the majority of the calculations reported here the atomic interactions
have been modeled using a non-orthogonal {\em tight-binding\/} scheme due to
Porezag and coworkers~\cite{porezag}. Tight-binding~(TB)
methods~\cite{tb:review} offer a good compromise between 
the more accurate but much more costly 
{\em first principles\/}~\cite{pw:review} techniques,
and {\em empirical potentials\/}~\cite{tersoff}, which 
are cheaper to use, but often not transferable to configurations different
to those for which they have been fitted.

In the TB scheme used here, 
the hopping integrals
used to construct the Hamiltonian and overlap matrices are tabulated as
a function of the internuclear distance on the basis of 
first-principles density-functional theory~(DFT) calculations employing
localized basis sets,
retaining only one- and two-center contributions to the Hamiltonian
matrix elements~\cite{porezag}. A minimal basis set corresponding to a 
single atomic-like orbital per atomic valence state is used. 
More details on the
construction of the TB parametrisation used in this work can be found
in ref.~\cite{porezag}. 

Using the non-orthogonal TB scheme briefly outlined above we have performed
a series of calculations aimed at characterizing the elastic properties of
single-wall nanotubes. In particular, we have considered C, BN,
$\mbox{BC}_3$ and $\mbox{BC}_2\mbox{N}$ (n,n) and (n,0) (i.e. non-chiral)
nanotubes. Two structures having the same stoichiometry are possible for the
$\mbox{BC}_2\mbox{N}$ nanotubes. 
Only the structure known as II~\cite{BCN,renata} is considered here, as this
is predicted to be the most stable of the two.
In addition, we have performed calculations for the
chiral (10,5) and (10,7) C nanotubes. 
We have also carried out Plane Wave~(PW)
pseudopotential DFT calculations within the local density approximation~(LDA)
for the (6,6) C and BN nanotubes,
for comparison with the TB results.  Our PW calculations
used Troullier-Martins~pseudopotentials~\cite{troullier:martins}. 
A cut off of 40~Ry was used in the
basis set, and 10~Monkhorst-Pack~\cite{pw:review} points to sample the
one-dimensional Brillouin zone. 
The hexagonal supercell was chosen
large enough so as to ensure that the minimum distance between a tube and 
any of its periodic images was larger than 5.5~\AA. 
Our TB calculations were performed using $\Gamma$-point sampling only, but
using periodically repeated cells which where large enough along the axial
direction so as to ensure that total energy differences were converged 
to an accuracy
approximately equal to that achieved with the PW calculations. 

The Poisson ratio is defined via the equation
\begin{eqnarray}
\frac{R - R_{eq}}{R_{eq}} = -\sigma \epsilon.
\end{eqnarray}
Here, $\epsilon$ is the axial strain, $R_{eq}$ is the equilibrium tube 
radius, $R$ is the tube radius at strain $\epsilon$ and $\sigma$ is 
Poisson's ratio.  The values of 
$\sigma$ obtained for a number of representative tubes considered in this work 
are reported in Table~\ref{table:tb_properties}.
Regarding the Young modulus, its conventional definition is 
\begin{eqnarray}
Y = \left. \frac{1}{V_0}\; \frac{\partial^2 \! E}{\partial\, \epsilon^2}
            \right|_{\epsilon=0},
\end{eqnarray}
where $V_0$ is the equilibrium volume, and $E$ is the strain energy. In the 
case of a single-wall nanotube, this definition requires adopting a convention 
in order to define $V_0$, which for a hollow cylinder is given by 
$V_0 = 2\pi L R \delta R$, where $L$ is the length, $R$ the radius and
$\delta R$ is the shell thickness. 
Different conventions have been adopted in the
past; for example, Lu~\cite{lu} recently took $\delta R = 0.34$~nm, i.e. the
interlayer separation in graphite, while Yakobson {\em et al\/}~\cite{yakobson}
took the value $\delta R = 0.066$~nm. We follow a different path. Rather
than adopting an {\em ad hoc\/} convention, we use a different magnitude
to characterize the stiffness of a single-wall nanotube, which is independent
of any shell thickness. We define
\begin{eqnarray}
Y_s = \left. \frac{1}{S_0} \; \frac{\partial^2\! E}{\partial\, \epsilon^2}
		\right|_{\epsilon=0}.
\end{eqnarray}
Here, $S_0$ is the surface defined by the tube at equilibrium. 
The value of the Young modulus for a given convention value $\delta R$ is
given by $Y = Y_s/\delta R$. In Table~\ref{table:tb_properties} we
give the values obtained for $Y_s$ for a number of tubes.

\begin{figure}
\begin{center}
\leavevmode
\epsfxsize=8cm
\epsffile{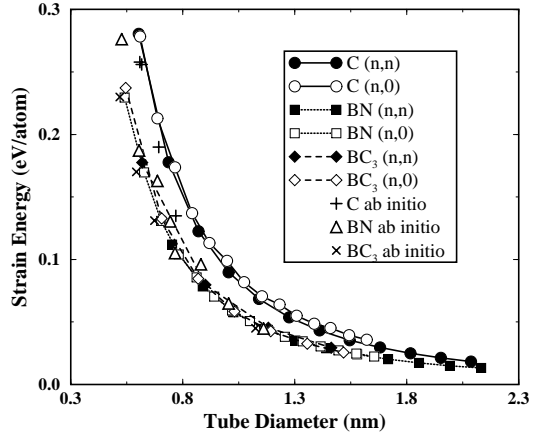}
\end{center}
\caption{Curvature strain energy as a function of the equilibrium tube
diameter, as obtained from the tight-binding calculations, for C, BN
and $\mbox{BC}_3$ nanotubes. {\em ab initio\/} results are from 
refs.~[8-10].}
\label{fig:curvature}
\end{figure}

Our first observation 
is that PW and TB results give very similar answers for all the calculated
properties. The differences in values of the Young modulus (calculated 
adopting the convention $\delta R = 0.34$~nm) for the (6,6) C nanotube are
of the order of 0.12~TPa. 
\begin{table}
\begin{tabular}{|c|ccccc|}
$\mbox{B}_x\mbox{C}_y\mbox{N}_z$ & (n,m) & $D_{eq}$~(nm) & $\sigma$ & 
$Y_s$~($\mbox{TPa}\cdot\mbox{nm}$) & $Y$~(TPa) \\
\hline
C & (10,0) & 0.791 & 0.275 & 0.416 & 1.22 \\
  & (6,6) & 0.820 & 0.247 & 0.415 & 1.22 \\
  &       & (0.817)      &       & (0.371) & (1.09) \\
  & (10,5) & 1.034 & 0.265 & 0.426 & 1.25 \\
  & (10,7) & 1.165 & 0.266 & 0.422 & 1.24 \\ 
  & (10,10) & 1.360 & 0.256 & 0.423 & 1.24 \\
  & (20,0) & 1.571 & 0.270 & 0.430 & 1.26 \\
  & (15,15) & 2.034 & 0.256 & 0.425 & 1.25 \\ \hline
BN & (10,0) & 0.811 & 0.232 & 0.284 & 0.837 \\
   & (6,6) & 0.838 & 0.268 & 0.296 & 0.870 \\
   &       & (0.823)      &       & (0.267) & (0.784) \\
   & (15,0) & 1.206 & 0.246 & 0.298 & 0.876 \\
   & (10,10) & 1.390 & 0.263 & 0.306 & 0.901 \\
   & (20,0) & 1.604 & 0.254 & 0.301 & 0.884 \\ 
   & (15,15) & 2.081 & 0.263 & 0.310 & 0.912 \\ \hline
$\mbox{BC}_3$ & (5,0) & 0.818 & 0.301 & 0.308 & 0.906 \\
              & (3,3) & 0.850 & 0.289 & 0.311 & 0.914 \\
              & (10,0) & 1.630 & 0.282 & 0.313 & 0.922 \\ 
              & (6,6) & 1.694 & 0.279 & 0.315 & 0.925 \\ \hline
$\mbox{BC}_2\mbox{N}$ II & (7,0) & 1.111 & 0.289 & 0.336 & 0.988 \\
                         & (5,5) & 1.370 & 0.287 & 0.343 & 1.008 \\
\end{tabular}
\caption{Structural and elastic properties of selected nanotubes obtained
from the tight-binding calculations reported here. Young modulus 
values given in parenthesis were obtained from first-principles calculations.
Also the value of $Y$ 
with the convention $\delta R = 0.34$~nm is given for comparison.} 
\label{table:tb_properties}
\end{table} 
It is worth pointing out that this difference is small compared
to the uncertainty with which $Y$ can at present be experimentally 
determined for multi-wall nanotubes~\cite{treacy,lieber,zettl}. 
Structural properties deduced from the PW and TB calculations are also in
very good agreement for both C and BN systems. The equilibrium diameter, 
as seen in 
Table~\ref{table:tb_properties}, differs by about 1~\%\ or less.
The nearest neighbor distance in the case of the (6,6) C nanotube
is 1.42~\AA\ in both the PW and TB calculations.
For the BN (6,6) nanotube, the results are 1.43~\AA\
and 1.45~\AA\ for the PW and TB calculations, respectively.

An interesting magnitude associated with nanotubes is the curvature energy
or {\em strain energy\/}~$E_s$, which we define as the difference of the energy
per atom in the tube and that in the corresponding infinite flat sheet. In 
Fig.~\ref{fig:curvature} we plot the strain energy obtained from our TB 
calculations for C, BN and $\mbox{BC}_3$ nanotubes as a function of the
tube diameter. It can be seen that the characteristic behavior
$E_s \propto 1/D^2$, where $D$ is the tube diameter, is obtained. Our 
calculations indicate that the strain energy at a given tube diameter is
highest for C nanotubes, and that both BN and $\mbox{BC}_3$ nanotubes have
very nearly the same strain energy. The fact that these composite nanotubes
have smaller strain energy than pure carbon nanotubes is in agreement with
previous first-principles results~\cite{rubio:corkill:cohen,miyamoto}.

\begin{figure}
\begin{center}
\leavevmode
\epsfxsize=8cm
\epsffile{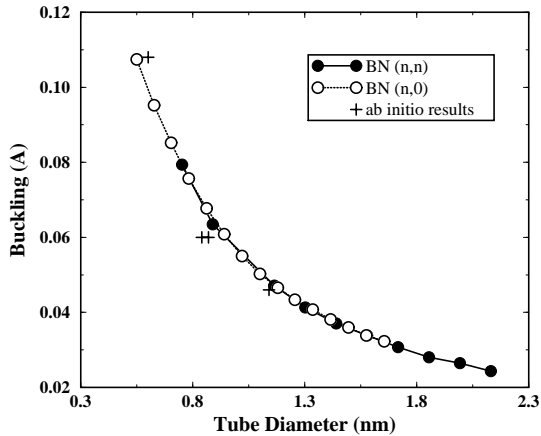}
\end{center}
\caption{Buckling in the BN nanotube equilibrium structures vs. 
tube diameter. We define the buckling as the mean radius of the 
nitrogen atoms minus the mean radius of the boron atoms. The 
{\em ab initio\/} results are from ref.~[8].}
\label{fig:bn_buckling}
\end{figure}

A structural feature which is specific to the BN nanotubes is the presence
of a certain degree of buckling on the tube surface, which results from the
B atoms displacing towards the tube axis, while the N atoms displace
in the opposite direction. 
Notice the good agreement with {\em ab initio\/} 
results~\cite{rubio:corkill:cohen}.
As for the strain energy, the buckling effect decreases rapidly with
increasing tube diameter, going to the flat BN sheet limit of zero buckling.
This tendency of BN nanotubes to buckle, which is a result of the 
slightly different hybridizations of B and N in the curved hexagonal layer,
will have the effect of forming a surface dipole, a fact that could be relevant
for potential applications of these tubes.

\begin{figure}
\begin{center}
\leavevmode
\epsfxsize=8cm
\epsffile{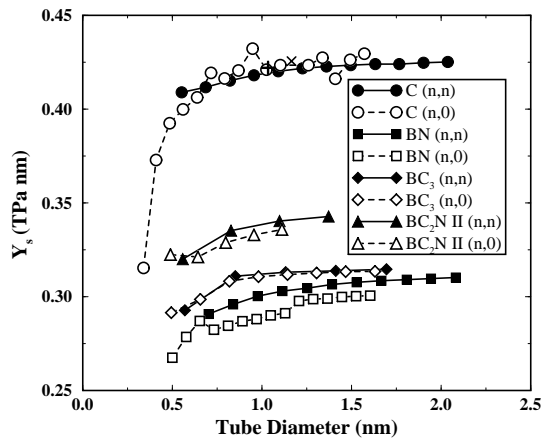}
\end{center}
\caption{Young modulus as a function of the tube diameter for C, BN,
$\mbox{BC}_3$ and $\mbox{BC}_2\mbox{N}$ (structure II only), as
calculated from the tight-binding simulations.
Results obtained for (n,n) nanotubes (filled symbols), (n,0) nanotubes
(empty symbols) and also for C (10,5) ($+$) and (10,7) ($\times$) are shown.}
\label{fig:modulus}
\end{figure}

In Fig.~\ref{fig:modulus} we have plotted the values of $Y_s$ obtained
for the different tubes. The first feature to be noticed is the fact that 
both for (n,n) and (n,0) nanotubes, the carbon tubes are predicted to
be significantly stiffer than any of the composite tubes. 
The $\mbox{BC}_2\mbox{N}$ are
predicted to be somewhat stiffer than BN and $\mbox{BC}_3$ tubes. 
The value of 0.43~$\mbox{TPa}\cdot \mbox{nm}$ obtained for the widest C nanotubes
corresponds to a Young modulus of 1.26~TPa, 
taking a value of 0.34~nm for the graphene sheet thickness. This value is
in excellent agreement with the experimental result of $1.28\pm 0.59$~TPa 
of Wong~{\em et al.\/}~\cite{lieber}.
Although our results are only for single-wall tubes, it can be expected that
the elastic properties of multi-wall tubes and nanoropes 
be mostly determined by the strength
of the C--C bonds in the bent graphene sheets, and thus be
very similar to those of single-wall tubes. 
Our results for C nanotubes are also in reasonable agreement
with the measurements of Treacy {\em et al.\/}~\cite{treacy}
($1.8\pm 1.4$~TPa).
Chopra and Zettl~\cite{zettl} 
obtain a value of 1.22~TPa with an
estimated 20~\%\ error for multi-wall BN nanotubes. 
This value is again somewhat larger than what we obtain
for BN nanotubes, but nevertheless, the agreement is close.
Lu's~\cite{lu}
estimation of the Young modulus for single-wall C nanotubes gives results
which are slightly smaller than ours (0.97~TPa), a difference which is most
likely due to the different models used in his calculations and ours.

Our calculations predict that there is a small dependence of $Y_s$ on the
tube diameter, but this dependence is noticeable only for small values of
the tube diameter, the limiting (diameter independent) value being rapidly
obtained at the range of experimentally observed single-wall tube diameters
($\sim 1.2$~nm). 
This is in contrast to the results of Lu~\cite{lu}, which are 
almost completely independent of the tube size and chirality. We believe 
that the apparent insensibility of the Young modulus on the tube size and 
chirality observed by Lu is due to the fact that an empirical pair potential
was used in his calculations, and such a model will not reflect the effects
that the curvature will have on the bonding properties of the system. 
In the limit of large tube diameters, we could expect that the
elastic properties would correspond to those of a plane, defect-free,
graphitic sheet.
Indeed, calculations of $Y_s$
for plane graphene and BN sheets give
0.41 and 0.30~$\mbox{TPa}\cdot\mbox{nm}$ respectively, 
which can be seen to be very
similar to the results obtained for C and BN nanotubes of the largest 
diameter we studied. It is worth noticing that the limiting value
of the Young modulus as a function of tube diameter is reached from 
below, which is consistent with the expectation that tubes of higher
curvature (i.e. smaller diameter) will have weaker bonds, which would
result in a slight reduction of the Young modulus. 

To summarize, we have carried out an extensive study of the energetic, 
structural and elastic properties of both graphitic and composite nanotubes,
using a non-orthogonal Tight-Binding scheme. 
The agreement
obtained between the TB results and the first-principles calculations 
reassure us in our conclusion that
the TB model employed here gives a good description of the studied features
of nanotubes. We have obtained good agreement with the existing experimental
measurements of the Young modulus for multi-wall C and BN nanotubes. Our 
results indicate that graphitic nanotubes are stiffer than any of the composite
nanotubes considered in this work, and that the elastic properties of 
single-wall nanotubes are of the same order of magnitude as those of the
corresponding flat sheets. Although the BN nanotubes are predicted 
to have a somewhat smaller Young modulus than the C nanotubes, they remain
considerably stiff. This fact, combined with their insulator 
character~\cite{rubio:corkill:cohen} makes them suitable for 
applications in which electrically insulating high-strength materials 
are needed. 

We would like to express our gratitude to G.~Seifert and T.~Heine for
providing the TB parameters used in this work and for helpful discussions
concerning implementation details. We are also grateful to
P.~Bernier and to M.~Galtier for
stimulating discussions. Financial support was provided by the
EU through its Training and Mobility of Researchers Programme under
contract ERBFMRX-CT96-0067 (D612-MITH). The use of computer facilities at C4
(Centre de Computaci\'{o} i Comunicacions de Catalunya) 
and CNUSC (Montpellier) is also acknowledged.

\end{document}